\documentclass[aps,twocolumn,showpacs,preprintnumbers,floats]{revtex4}\def\@cite#1#2{\textsuperscript{[{#1\if@tempswa , #2\fi}]}}
\usepackage{mathrsfs}
\usepackage{longtable,lscape}
\usepackage{txfonts}
\usepackage{amssymb}
\usepackage{indentfirst}
\usepackage{graphicx,booktabs}
\usepackage{multirow}
\usepackage{morefloats}
\usepackage{color}
\usepackage{graphicx}
\usepackage{amsfonts}
\usepackage{subfigure}

\usepackage{epsfig}
\usepackage[colorlinks, citecolor=blue,anchorcolor=red,menucolor=red, linkcolor=red,filecolor=red,urlcolor=blue,frenchlinks=red]{hyperref}

\newcommand{\vlab}{\mbox{\boldmath$\lambda$\unboldmath}}

\newcommand{\vxi}{\mbox{\boldmath$\xi$\unboldmath}}

\begin{document}

%\begin{spacing}{2.0}

\title{Charmed-strange tetraquarks and their decays in a potential quark model}
\author{Feng-Xiao Liu$^{1,4}$, Ru-Hui Ni$^{1,4}$, Xian-Hui Zhong$^{1,4}$~\footnote {E-mail: zhongxh@hunnu.edu.cn}, Qiang Zhao$^{2,3,4}$~\footnote {E-mail: zhaoq@ihep.ac.cn}}

\affiliation{ 1) Department of Physics, Hunan Normal University, and Key Laboratory of Low-Dimensional Quantum Structures and Quantum Control of Ministry of Education, Changsha 410081, China }

\affiliation{ 2) Institute of High Energy Physics, Chinese Academy of Sciences, Beijing 100049, China}

\affiliation{ 3) University of Chinese Academy of Sciences, Beijing 100049, China}

\affiliation{ 4)  Synergetic Innovation Center for Quantum Effects and Applications (SICQEA),
Hunan Normal University, Changsha 410081, China}
%

%\date{\today}

\begin{abstract}

In the framework of a nonrelativistic potential quark model, we investigate the mass spectrum of the $1S$-wave charmed-strange tetraquark states of $cn\bar{s}\bar{n}$ and $cs\bar{n}\bar{n}$ ($n=u$ or $d$) systems. The tetraquark system is solved by a correlated Gaussian method. With the same parameters fixed by the meson spectra, we obtained the mass spectra for the $1S$-wave tetraquark states. Furthermore, based on the predicted tetraquark spectra we estimate their rearrangement decays in a quark-exchange model. We find that the rearrangement decays of the tetraquarks may be mainly driven by the spin-spin interactions. The resonances $X_0(2900)^0$ and $T^a_{c\bar{s}0}(2900)^{++/0}$ reported from LHCb may be assigned to be the lowest $1S$-wave tetraquark states $\bar{T}_{cs0}^f(2818)$ and $T^{a}_{c\bar{s}0}(2828)$ classified in the quark model, respectively. It also allows us to extract the couplings for the initial tetraquark states to their nearby $S$-wave interaction channels. We find that some of these couplings turn out to be sizeable. Following the picture of the wavefunction renormalization for the near-threshold strong $S$-wave interactions, the sizeable coupling strengths can be regarded as an indication of their dynamic origins as candidates for hadronic molecules. Furthermore, our predictions suggest that signals for the $1S$-wave charmed-strange tetraquark states can also be searched in the other channels, such as $D^0K^+$, $D^+K^+$, $D^{*+}K^-$, $D^{*+}K^+$, $D^{*0}K^+$, $D^0\bar{K}^{*0}$, $D_s^+\rho^0$, etc.

\end{abstract}


\maketitle

\section{Introduction}\label{indu}

Searching for genuine exotic hadrons beyond the conventional quark model has been one of the most important initiatives since the establishment of nonrelativistic constituent quark model (NRCQM) in 1964~\cite{GellMann:1964nj,Zweig:1981pd}.
Benefited from great progresses in experiment, strong evidences for exotic hadrons have been collected since the discovery of $X(3872)$ by Belle in 2003~\cite{Choi:2003ue}. Recent reviews of the status of experimental and theoretical studies can be found in Refs.~\cite{Liu:2019zoy,Esposito:2016noz,
Olsen:2017bmm,Lebed:2016hpi,Chen:2016qju,Ali:2017jda,Guo:2017jvc,Chen:2022asf}.
While many observed candidates have been found to be located in the vicinity of $S$-wave open thresholds, no signals for overall-color-singlet multiquark states have been indisputably established due to difficulties of distinguishing them from hadronic molecules~\cite{Guo:2017jvc}.

Very recently, the LHCb Collaboration reported their preliminary results on the observations of $cq\bar{s}\bar{q}$ tetraquarks~\cite{LHCb:2022}.
Two new tetraquark candidates $T^a_{c\bar{s}0}(2900)^{++}$ and $T^a_{c\bar{s}0}(2900)^{0}$ were observed in the $D_{s}^{+}\pi^+$ and
$D_{s}^{+}\pi^-$ invariant mass spectra in two $B$-decay processes $B^+\to D^- D_s^+\pi^+$ and $B^0\to \bar{D}^0 D_s^+\pi^-$, respectively.
The isospin and spin-parity quantum numbers are determined to be $(I)J^P=(1) 0^+$.
These two states should correspond to the two different charged states of the isospin triplet.
The measured mass and width are $M_{exp}=2908\pm 11\pm 20$ MeV and $\Gamma_{exp}=136\pm 23\pm 11$ MeV.
The $T^a_{c\bar{s}0}(2900)$ may be a flavor partner of the $0^+$ state $X_0(2900)$ (composed [$\bar{c}\bar{s}ud$]) observed in the $D^-K^+$ final state in $B^+\to D^+D^-K^+$ at LHCb in 2020~\cite{LHCb:2020bls,LHCb:2020pxc}.
The least quark components for $T^a_{c\bar{s}0}(2900)^{++}$, $T^a_{c\bar{s}0}(2900)^{0}$, and $X_0(2900)$ are $cu\bar{s}\bar{d}$, $cd\bar{s}\bar{u}$, and $\bar{c}\bar{s}ud$, respectively. Thus, from the quark contents, these states are ideal candidates of the exotic charmed-strange tetraquarks.

Relevant theoretical studies of the charmed-strange tetraquarks can be found in the literature~\cite{Chen:2017rhl,Agaev:2017oay,Zhang:2018mnm,Zhang:2020oze,Lu:2020qmp,Cheng:2020nho,
Wang:2020prk,Tan:2020cpu,Mutuk:2020igv,Wang:2020xyc,Chen:2020aos,He:2020jna,Karliner:2020vsi,Albuquerque:2021svg,
Guo:2021mja,Yang:2021izl,Agaev:2021jsz,Sundu:2022kyd,Agaev:2022eeh,Wei:2022wtr,Ke:2022ocs,Molina:2022jcd,Yang:2023evp,Lian:2023cgs,Agaev:2022eyk,Yue:2022mnf,Agaev:2022duz,
Dmitrasinovic:2023eei,Dmitrasinovic:2004cu,Dmitrasinovic:2005gc,Liu:2020orv,Burns:2020epm,Agaev:2020nrc,Wang:2021lwy,He:2020btl,Xiao:2020ltm,
Molina:2020hde,Huang:2020ptc,Xue:2020vtq,Liu:2020nil}, among which most of these works were stimulated by the discovery of
$X_0(2900)$ and $T^a_{c\bar{s}0}(2900)$. It should be mentioned that for $X_0(2900)$
and/or $T^a_{c\bar{s}0}(2900)$, apart from the compact tetraquark interpretation~\cite{Mutuk:2020igv,Yang:2021izl,Tan:2020cpu,Yang:2023evp,Dmitrasinovic:2023eei,Lian:2023cgs,Wei:2022wtr,Albuquerque:2021svg,
Wang:2020prk,Wang:2020xyc,He:2020jna,Karliner:2020vsi,Guo:2021mja},
there are also other possible interpretations, such as
hadronic molecule states~\cite{Chen:2020aos,Agaev:2022eeh,Ke:2022ocs,Liu:2020nil,Huang:2020ptc,Molina:2020hde,Xue:2020vtq,Agaev:2020nrc,
Xiao:2020ltm,He:2020btl,Wang:2021lwy,Agaev:2022eyk,Yue:2022mnf,Agaev:2022duz}, and threshold effects~\cite{Liu:2020orv,Burns:2020epm,Molina:2022jcd}. In particular, for the exotic candidate observed in $D_{s}^{+}\pi^+$ it is inevitable that its overall color-singlet configuration would couple to those allowed two-body thresholds. If the physical state is close to the nearby $S$-wave threshold and has strong couplings, it implies that there should exist a sizeable hadronic molecular component within the exotic candidate as the long-range component of the wavefunction. Meanwhile, the short-range component should be driven by the non-perturbative dynamics among the constituent quarks as a tetraquark~\cite{Weinberg:1962hj,Weinberg:1963zza,Weinberg:1965zz,Guo:2014iya}. This makes it interesting to study the four-body constituent quark system in the quark model and investigate the decays of the tetraquark states into the nearby two-body channels.

In this work, to understand the nature of the newly observed exotic resonances $T^a_{c\bar{s}0}(2900)$ and $X_0(2900)$,
we carry out a systematic study of the mass spectrum of the $1S$-wave charmed-strange tetraquarks in a nonrelativistic potential quark model (NRPQM). The NRPQM is based on the Hamiltonian of the Cornell model~\cite{Eichten:1978tg}, which has made great successes in the description of the charmonium and bottomonium spectra with high precision, and been broadly applied to multiquark systems in the literature. It contains a linear confinement and a one-gluon-exchange (OGE) potential for quark-quark and quark-antiquark interactions. To solve the four-body problem accurately, the explicitly correlated Gaussian method is adopted in our calculations.
Furthermore, we have analyzed the rearrangement decays of the $1S$-states in a
quark-exchange model. The transition operators can be extracted from the quark-quark and quark-antiquark interactions in the NRPQM. This guarantees a self-consistent treatment of the eigenstates and their decays. This will allow a better understanding of the dynamic origin of the tetraquark candidates $T^a_{c\bar{s}0}(2900)^{++,0}$ and $X_0(2900)$ and their couplings to the continuum states.
%of the LHCb observations can fit in the compact $1S$-wave charmed-strange tetraquarks obtained by a NRPQM.

As follows, we first give a brief introduction to our framework. We then present the full numerical results
for the $S$-wave charmed-strange tetraquark states to compare with the experimental observations.
Phenomenological consequence and implications for future experimental studies will be discussed.

%\begin{figure}[]
%\centering \epsfxsize=6.6 cm \epsfbox{csqq.eps} \vspace{-0.5cm} \caption{Mass spectra for the $cc\bar{c}\bar{c}$ and $bb\bar{b}\bar{b}$ systems.}\label{figmasss}
%\end{figure}

%{\emph{Model and method}}--
%

\section{Framework}

\subsection{Mass spectrum}\label{DISSCUS}

\subsubsection{Hamiltonian}

The mass spectrum of the tetraquarks are calculated within the NRPQM, which has been widely adopted to deal with the mass spectra of mesons and baryons. In this model the Hamiltonian is given by~\cite{Liu:2021rtn,Liu:2019zuc,Liu:2020lpw}
\begin{equation}\label{Hamiltonian}
H=(\sum_{i=1}^4 m_i+T_i)-T_G+\sum_{i<j}V_{ij}(r_{ij}),
\end{equation}
where $m_i$ and $T_i$ stand for the constituent quark mass and kinetic energy of the $i$th quark, respectively; $T_G$ stands for the center-of-mass (c.m.) kinetic energy of the tetraquark system; $r_{ij}\equiv|\mathbf{r}_i-\mathbf{r}_j|$ is the distance between the $i$th and  $j$th quark. The two-body effective potentials between quarks, $V_{ij}(r_{ij})$, are given by
\begin{eqnarray}\label{vcon}
V_{ij}(r_{ij})&&=-\frac{3}{16}({\vlab}_i\cdot{\vlab}_j)\left( b_{ij}r_{ij}-\frac{4}{3}\frac{\alpha_{ij}}{r_{ij}}+C_0 \right)\nonumber\\
&&-\frac{\alpha_{ij}}{4}({\vlab}_i\cdot{\vlab}_j)\left\{\frac{\pi}{2}\cdot\frac{\sigma^3_{ij}e^{-\sigma^2_{ij}r_{ij}^2}}{\pi^{3/2}}
\cdot\frac{16}{3m_im_j}(\mathbf{S}_i\cdot\mathbf{S}_j)\right\},
\end{eqnarray}
where the first term is the confinement potential part, which adopts the standard form of the Cornell potential~\cite{Eichten:1978tg};  while the second term is the spin-spin potential part. In the above equation, the constant $C_0$ stands for the zero point energy; $\mathbf{S}_i$ stands for the spin of the $i$th quark, $\vlab_i$ are the color generators of SU(3) group; The parameters $b_{ij}$ and $\alpha_{ij}$ denote
the strength of the confinement and strong coupling of the one-gluon-exchange potential, respectively.
It should be mentioned that the tensor and spin-orbit potential do not contribute to the $1S$-wave tetraquarks considered here .

The model parameters $m_{i,j}$, $\alpha_{i j}$, $b_{i j}$, $\sigma_{i j}$ and $C_{0}$
adopted in this work have been listed in Table~\ref{para}, which are extracted by fitting the mass
spectra of the non-strange, strange, charmed and charm-strange mesons~\cite{ParticleDataGroup:2020ssz} as also shown in Table~\ref{para}.

The NRPQM not only gives successful descriptions of the $b\bar{b}$ and $c\bar{c}$ states,
but also obtains acceptable results for the meson spectra containing clearly
relativistic light quarks~\cite{Vijande:2004he,Lucha:1991vn}.
Several studies in the literature~\cite{Lucha:1989jf,Semay:1992xq,Jaczko:1998uj} have been carried out
to understand why an ostensibly nonrelativistic treatment works and allows
useful predictions to be made for relativistic systems.
For a heavy quark with mass of $m$ and three-momentum $\mathbf{p}$,
the relativistic kinetic term can be expanded with the standard expansion in powers of $p^2/m^2$, i.e.,
\begin{equation}
\sqrt{m^2+p^2}=m+\frac{p^2}{2m}-\frac{p^4}{8m^3}+\cdot\cdot\cdot.
\end{equation}
However, this expansion fails for a light quark. A possible solution to this
problem is to consider an expansion about a fixed momentum $p^2_0$~\cite{Jaczko:1998uj},
\begin{eqnarray}
\sqrt{m^2+p^2}&=&\sqrt{p^2-p_0^2+M^2}\nonumber\\
&=&M+\frac{p^2-p_0^2}{2M}-\frac{(p^2-p_0^2)^2}{8M^3}+\cdot\cdot\cdot,
\end{eqnarray}
where $M=\sqrt{m^2+p_0^2}$ can be considered as an effective quark mass. The expansion will give a good average
approximation to the relativistic kinetic energy provided
the relevant values of $p^2$ are concentrated near $p^2_0$ with
$\langle(p^2-p_0^2)^2\rangle\ll M^4$. Taking $p^2_0=\langle(p^2)\rangle$, the
relativistic kinetic term for a light quark can be approximated as
\begin{eqnarray}\label{nonH}
\sqrt{m^2+p^2}\simeq M+\frac{p^2}{2M}+\epsilon,
\end{eqnarray}
with $\epsilon=-\langle p^2\rangle/(2M)-\langle p^2-\langle p^2\rangle \rangle/(8M)$.
In some case the $\epsilon$ term can be approximately considered as a constant
term, which can be absorbed in the zero point energy parameter $C_0$ of the potential.
Thus, from Eq.~(\ref{nonH}) one finds that the kinetic term for a light quark still can be
expressed as the often used nonrelativistic form, the relativistic effects
are absorbed in the parameters of constituent quark mass and zero
point energy. Since there is good equivalence between relativistic and
nonrelativistic quark models, in this work we adopt the nonrelativistic
quark model, with which our calculations for the tetraquarks become more easy
than that with relativistic models.

\begin{table}
\begin{center}
\caption{The quark model parameters determined by fitting the meson mass spectra. The unit of the meson masses is MeV.}\label{para}
\scalebox{1.0}{
\begin{tabular}{llllllllll}
\hline
\hline
&\multicolumn{2}{l}{\underline{~~~~~~~~~~~~~Parameter set~~~~~~~~~~~~~~}}
&\multicolumn{3}{c}{\underline{~~~~Meson mass spectrum~~~~} }\\
&  &  &State & Ours   &Exp.~\cite{ParticleDataGroup:2020ssz} \\
\hline
&$m_{u/d}$ [GeV]                       &$0.35$          ~~~&$\pi$             ~~~&$135$   ~~~~~&$135$       \\
&$m_{s}$ [GeV]                       &$0.5$           ~~~&$\rho(770)$       ~~~&$775$   ~~~~~&$775$       \\
&$m_{c}$ [GeV]                       &$1.5$           ~~~&$a_2(1320)$       ~~~&$1305$  ~~~~~&$1318$      \\
&$\alpha_{nn},\alpha_{sn}$           &$0.990$         ~~~&$\rho_3(1690)$    ~~~&$1637$  ~~~~~&$1689$      \\
&$\alpha_{cn},\alpha_{cs}$           &$0.635$         ~~~&$K$               ~~~&$498$   ~~~~~&$498$       \\
&$b_{nn},b_{sn}$ $[\mathrm{GeV}^2]$  &$0.140$         ~~~&$K^*(892)$        ~~~&$892$   ~~~~~&$892$       \\
&$b_{cn},b_{cs}$ $[\mathrm{GeV}^2]$  &$0.140$         ~~~&$K_2^*(1430)$     ~~~&$1457$  ~~~~~&$1427$      \\
&$\sigma_{nn}$ [GeV]                 &$0.574$   ~~~&$K_3^*(1780)$     ~~~&$1785$  ~~~~~&$1779$  \\
&$\sigma_{sn}$ [GeV]                 &$0.506$  ~~~&$D$               ~~~&$1865$  ~~~~~&$1865$      \\
&$\sigma_{cn}$ [GeV]                 & $0.787$        ~~~&$D^*(2007)$       ~~~&$2008$  ~~~~~&$2008$      \\
&$\sigma_{cs}$ [GeV]                 & $0.831$        ~~~&$D_2^*(2460)$     ~~~&$2454$  ~~~~~&$2461$      \\
&$C_{0}(nn)$ [MeV]                   & $-456.0$       ~~~&$D_3^*(2750)$     ~~~&$2746$  ~~~~~&$2763$      \\
&$C_{0}(sn)$ [MeV]                  &$-380.0$        ~~~&$D_s$             ~~~&$1969$  ~~~~~&$1969$      \\
&$C_{0}(cn)$ [MeV]                   &$-286.0$        ~~~&$D_s^*$           ~~~&$2112$  ~~~~~&$2112$      \\
&$C_{0}(cs)$ [MeV]                   &$-220.0$        ~~~&$D_{s2}(2573)$    ~~~&$2573$  ~~~~~&$2569$      \\
&                                    &                ~~~&$D_{s3}^*(2860)$  ~~~&$2861$  ~~~~~&$2860$  \\
\hline
\hline
\end{tabular}}
\end{center}
\end{table}

\begin{table}
\begin{center}
\caption{Flavor wave functions of the tetraquark systems $cn\bar{s}\bar{n}$ and $cs\bar{n}\bar{n}$.
In the table we define $\{\bar{q}_3\bar{q}_4\}=\sqrt{\frac{1}{2}}(\bar{q}_3\bar{q}_4+\bar{q}_4\bar{q}_3)$ and $[\bar{q}_3\bar{q}_4]=\sqrt{\frac{1}{2}}(\bar{q}_3\bar{q}_4-\bar{q}_4\bar{q}_3)$.}\label{flavor}
\begin{tabular}{cccccccccc}
\hline\hline
& $I$& $I_3$&$\underline{~~~~~~~~~~6_F~~~~~~}$&&$\underline{~~~~~~~~\bar{3}_F~~~~~~~}$\\
\hline
\multirow{4}{*}{$cn\bar{s}\bar{n}$}
&$0$&$0$& $\sqrt{\frac{1}{2}}cd\{\bar{s}\bar{d}\}+\sqrt{\frac{1}{2}}cu\{\bar{s}\bar{u}\}$ &  & $\sqrt{\frac{1}{2}}cd[\bar{s}\bar{d}]+\sqrt{\frac{1}{2}}cu[\bar{s}\bar{u}]$  \\
&   1    &$+1$& $cu\{\bar{s}\bar{d}\}$ &  & $cu[\bar{s}\bar{d}]$  \\
&   1    &$0$& $\sqrt{\frac{1}{2}}cd\{\bar{s}\bar{d}\}-\sqrt{\frac{1}{2}}cu\{\bar{s}\bar{u}\}$ &  & $\sqrt{\frac{1}{2}}cd[\bar{s}\bar{d}]-\sqrt{\frac{1}{2}}cu[\bar{s}\bar{u}]$  \\
&   1    &$-1$& $cd\{\bar{s}\bar{u}\}$ &  & $cd[\bar{s}\bar{u}]$  \\
\hline
\multirow{4}{*}{$cs\bar{n}\bar{n}$}
&   0     &$0$&      &  & $cs[\bar{u}\bar{d}]$  \\
&   1   &$+1$& $cs\bar{d}\bar{d}$ &  &   \\
&   1   &$0$& $cs\{\bar{u}\bar{d}\}$ &  &   \\
&   1   &$-1$& $cs\bar{u}\bar{u}$ &  &   \\
\hline
\hline
\end{tabular}
\end{center}
\end{table}

\subsubsection{Tetraquark configurations}

For a tetraquark system $Q_1q_2\bar{q}_3\bar{q}_4$ containing a heavy quark $Q$ and three
light quarks ($u,d$, or $s$), the $\bar{q}_3\bar{q}_4$ anti-quark pair should satisfy the SU(3) flavor symmetry.
As the result, the $Q_1q_2\bar{q}_3\bar{q}_4$ system can form two different SU(3) flavor
representations: the symmetric sextet $6_F$ and antisymmetric
antitriplet $\bar{3}_F$. By combining the SU(3) flavor symmetry
and the requirement of the isospin, one can obtain the flavor wave functions of
the tetraquark system $Q_1q_2\bar{q}_3\bar{q}_4$. In this work, we focus on the charmed-strange systems
$cn\bar{s}\bar{n}$ and $cs\bar{n}\bar{n}$ ($n=u$ or $d$),
whose flavor functions are explicitly given in Table~\ref{flavor}.
For simplicity, here we do not explicitly give the color wave functions and spin wave
functions, which can be found in our previous works~\cite{Liu:2021rtn,Liu:2019zuc,Liu:2020lpw}.

Considering the Pauli principle and color confinement for the $cn\bar{s}\bar{n}$ system,
we have 12 $1S$ configurations for $I=0$ and $I=1$, respectively,
while for the $cs\bar{n}\bar{n}$ we have 6 $1S$ configurations for $I=0$ and $I=1$, respectively. They are listed in Table~\ref{abcd}. The subscripts and superscripts are the spin quantum numbers and representations of the color SU(3) group, respectively. A symmetric spatial wave function is implied for the ground state.

\subsubsection{Numerical method}

To solve the four-body problem accurately, we adopt the explicitly correlated Gaussian method~\cite{Mitroy:2013eom,Varga:1995dm}.
It is a well-established variational method to solve quantum-mechanical few-body
problems in molecular, atomic, and nuclear physics.
For a tetraquark system $Q_1q_2\bar{q}_3\bar{q}_4$
with zero angular momentum, the coordinate part of the wave function is expanded in terms
of correlated Gaussian basis. Such a basis function can be written as
\begin{equation}\label{spatial function1}
\psi=\exp\left[-\sum_{i<j=1}^4 \frac{1}{2a_{ij}^2}(\mathbf{r}_i-\mathbf{r}_j)^2\right],
\end{equation}
where $a_{ij}$ are adjustable parameters.
Considering the light anti-quark pair $\bar{q}_3\bar{q}_4$ as an identical
particle system, we can take $a_{13}=a_{14}\equiv c$ and $a_{23}=a_{24}\equiv d$
in the SU(3) symmetry limit. It is convenient to use a set of the Jacobi coordinates $\vxi=(\vxi_1, \vxi_1,\vxi_3)$,
instead of the relative distance vectors $(\mathbf{r}_i-\mathbf{r}_j)$.
Then the correlated Gaussian basis function can be rewritten as
\begin{equation}\label{spatial function1}
G(\vxi,A)=\exp\left(-\sum_{i,j} A_{ij}\vxi_i\cdot\vxi_j\right)\equiv\exp\left(-\tilde{\vxi}A\vxi\right),
\end{equation}
where the Jacobi coordinates $\vxi=(\vxi_1, \vxi_2,\vxi_3)$ are defined by
 \begin{eqnarray}
    \cases{
      \vxi_1\equiv\mathbf{r_1}-\mathbf{r_2}\cr
      \vxi_2\equiv\mathbf{r_3}-\mathbf{r_4}\cr
      \vxi_3\equiv\frac{m_1\mathbf{r_1}+m_2\mathbf{r_2}}{m_1+m_2}-\frac{m_3\mathbf{r_3}+m_4\mathbf{r_4}}{m_3+m_4}},
     \end{eqnarray}
and $A$ is a $3\times 3$ symmetric positive-definite matrix
whose elements are variational parameters. It should be pointed out that
in matrix $A$ there are only four independent variational parameters
$\{g\equiv a_{12}, f\equiv a_{34}, c,d\}$.

The coordinate part of the trial wave function $\Psi(\vxi, A)$ can be formed as a linear
combination of correlated Gaussians
\begin{eqnarray}
\Psi(\vxi, A)=\sum_{k=1}^{\mathcal{N}} c_k G(\vxi,A_k).
\end{eqnarray}
The accuracy of the trial function depends on the length of the expansion $\mathcal{N}$ and the nonlinear
parameters $A_k$. In our calculations, following the method of Ref.~\cite{Hiyama:2003cu}, we let the
variational parameters form a geometric progression. For example,
for a variational parameter $d$, we take
\begin{eqnarray}
d_n=d_1 a^{n-1} (n=1,\cdot\cdot\cdot,n^{max}_d).
\end{eqnarray}
There are three parameters
$\{d_{1}, d_{n^{max}_d}, n^{max}\}$ to be determined through the variation method.
The length of the expansion $\mathcal{N}$ is determined to by
$\mathcal{N}=n^{max}_{g} n^{max}_{f} n^{max}_c n^{max}_d$.
In this work, we take $n^{max}_{g}=n^{max}_{f}=n^{max}_c=n^{max}_d=4$,
then we obtain stable solutions.

\subsection{Rearrangement decay}\label{DISSCUS}

With the eigenstates obtained in the previous section,
we may estimate the rearrangement decays of the $cs\bar{n}\bar{n}$ and $cn\bar{s}\bar{n}$ systems in a
quark-exchange model~\cite{Barnes:2000hu}. The transition operators are extracted from the quark-quark and
quark-antiquark interactions via the quark rearrangement. The decay amplitude $\mathcal{M}(A\to BC)$ of a tetraquark state is described by
\begin{eqnarray}
\mathcal{M}(A\to BC)=-\sqrt{(2\pi)^3}\sqrt{8M_AE_BE_C}\left\langle BC |\sum_{i<j} V_{ij}| A \right\rangle,
\end{eqnarray}
where $A$ stands for the initial tetraquark state, $BC$ stands for the final hadron pair.
$V_{ij}$ are the potentials between inner quarks of final hadrons $B$ and $C$, they are taken the same
as that of the potential model given in Eq.~(\ref{vcon}). $M_A$ is the mass of the
initial state, while $E_B$ and $E_C$ are the energies of the final states $B$ and $C$, respectively, in the initial-hadron-rest system.

This phenomenological model has been applied to the study of the hidden-charm decay properties for the multiquark states in the literature~\cite{Wang:2019spc,Xiao:2019spy,Wang:2020prk,Han:2022fup}.
For simplicity, the wave functions of the initial and final state hadrons, i.e. $A,\ B, \ C$, are adopted in the form of  single harmonic oscillator.  They are determined by fitting the wave functions calculated from our potential model.
The partial decay width of $A\to BC$ is given by
\begin{eqnarray}
\Gamma=\frac{1}{2J_A+1}\frac{|\mathbf{q}|}{8\pi M_A^2}\left|\mathcal{M}(A\to BC)\right|^2,
\end{eqnarray}
where $\mathbf{q}$ is the three-vector momentum of the final state $B$ or $C$ in the initial-hadron-rest
frame.

\begin{table}
\begin{center}
\caption{ The $1S$ configurations and the average contributions of each part of the Hamiltonian for the $cn\bar{s}\bar{n}$ and $cs\bar{n}\bar{n}$ systems. The unit is MeV. In the table, for the $I=0$ configurations we define that $cn\{\bar{s}\bar{n}\}\equiv (cd\{\bar{s}\bar{d}\}+cu\{\bar{s}\bar{u}\})/\sqrt{2}$, and $cn[\bar{s}\bar{n}]\equiv(cd[\bar{s}\bar{d}]+cu[\bar{s}\bar{u}])/\sqrt{2}$; while for the $I=1$ configurations, we define that $cn\{\bar{s}\bar{n}\}\equiv \{cu\{\bar{s}\bar{d}\}, (cd\{\bar{s}\bar{d}\}-cu\{\bar{s}\bar{u}\})/\sqrt{2}, cd\{\bar{s}\bar{u}\}$\} and $cq[\bar{s}\bar{q}]\equiv \{cu[\bar{s}\bar{d}], (cd[\bar{s}\bar{d}]-cu[\bar{s}\bar{u}])/\sqrt{2}, cd[\bar{s}\bar{u}]$\}. }\label{abcd}
\begin{tabular}{clccccc}
\hline\hline
\multicolumn{7}{l}{$cn\bar{s}\bar{n}$ system}\\
\hline
$(I)~J^{PC}$  & Configuration  & Mass  & $\langle T\rangle$  & $\langle V^{Lin}\rangle$  & $\langle V^{Coul}\rangle$  & $\langle V^{SS}\rangle$\\
\hline
\multirow{4}{*}{$(0,1)~0^{+}$}
 & $\left| \left(cn\right)_{0}^{6}\left\{ \bar{s}\bar{n}\right\} _{0}^{\bar{6}}\right\rangle _{0}$  & 3181&	 788&	 1054&	-723&	33\\
 & $\left| \left(cn\right)_{1}^{\bar{3}}\left\{ \bar{s}\bar{n}\right\} _{1}^{3}\right\rangle _{0}$  & 3145&	 877&	 1022&	-751&	-34\\
 & $\left| \left(cn\right)_{0}^{\bar{3}}\left[\bar{s}\bar{n}\right]_{0}^{3}\right\rangle_{0}    $  & 3053&	 984&	 966&	-802&	-127\\
 & $\left| \left(cn\right)_{1}^{6}\left[\bar{s}\bar{n}\right]_{1}^{\bar{6}}\right\rangle_{0}    $  & 2925&	 1118&	 889&	-859&	-250\\
\hline
\multirow{6}{*}{$(0,1)~1^{+}$}
 & $\left| \left(cn\right)_{1}^{6}\left\{ \bar{s}\bar{n}\right\} _{0}^{\bar{6}}\right\rangle_{1}$  & 3167&	 920&	 977&	-781&	23\\
 & $\left| \left(cn\right)_{0}^{\bar{3}}\left\{ \bar{s}\bar{n}\right\} _{1}^{3}\right\rangle_{1}$  & 3168&	 888&	 1016&	-759&	-9\\
 & $\left| \left(cn\right)_{1}^{\bar{3}}\left\{ \bar{s}\bar{n}\right\} _{1}^{3}\right\rangle _{1}$  & 3178&	 861&	 1031&	-745&	0\\
 & $\left| \left(cn\right)_{1}^{\bar{3}}\left[\bar{s}\bar{n}\right]_{0}^{3}\right\rangle_{1}    $  & 3093&	 933&	 994&	-779&	-85\\
 & $\left| \left(cn\right)_{0}^{6}\left[\bar{s}\bar{n}\right]_{1}^{\bar{6}}\right\rangle_{1}    $  & 3144&	 933&	 970&	-786&	0\\
 & $\left| \left(cn\right)_{1}^{6}\left[\bar{s}\bar{n}\right]_{1}^{\bar{6}}\right\rangle_{1}    $  & 3030&	 1001&	 934&	-817&	-117\\
\hline
\multirow{2}{*}{$(0,1)~2^{+}$}
 & $\left| \left(cn\right)_{1}^{\bar{3}}\left\{ \bar{s}\bar{n}\right\} _{1}^{3}\right\rangle_{2}$  & 3239&	 819&	 1055&	-729&	63\\
 & $\left|\left(cn\right)_{1}^{6}\left[\bar{s}\bar{n}\right]_{1}^{\bar{6}}\right\rangle_{2}    $  & 3214&	 775&	 1063&	-716&	63\\
\hline
\hline
\multicolumn{7}{l}{$cs\bar{n}\bar{n}$ system}\\
\hline
$(I)~J^{PC}$  & Configuration  & Mass  & $\langle T\rangle$  & $\langle V^{Lin}\rangle$  & $\langle V^{Coul}\rangle$  & $\langle V^{SS}\rangle$\\
\hline
\multirow{2}{*}{$(1)~0^{+}$}
 & $\left| \left(cs\right)_{0}^{6}\left\{ \bar{u}\bar{d}\right\} _{0}^{\bar{6}}\right\rangle_{0}$  & 3170&	 799&	 1035&	-739&	38\\
 & $\left| \left(cs\right)_{1}^{\bar{3}}\left\{ \bar{u}\bar{d}\right\} _{1}^{3}\right\rangle _{0}$  & 3136&	 891&	 1003&	-767&	-21\\
\multirow{2}{*}{$(0)~0^{+}$}
 & $\left| \left(cs\right)_{0}^{\bar{3}}\left[\bar{u}\bar{d}\right]_{0}^{3}\right\rangle_{0}    $  & 2992&	 1017&	 941&	-821&	-174\\
 & $\left| \left(cs\right)_{1}^{6}\left[\bar{u}\bar{d}\right]_{1}^{\bar{6}}\right\rangle _{0}    $  & 2921&	 1086&	 887&	-863&	-226\\
\hline
\multirow{3}{*}{$(1)~1^{+}$}
 & $\left| \left(cs\right)_{1}^{6}\left\{ \bar{u}\bar{d}\right\} _{0}^{\bar{6}}\right\rangle_{1}$  & 3159&	 812&	 1026&	-745&	29\\
 & $\left| \left(cs\right)_{0}^{\bar{3}}\left\{ \bar{u}\bar{d}\right\} _{1}^{3}\right\rangle_{1}$  & 3165&	 870&	 1016&	-757&	6\\
 & $\left| \left(cs\right)_{1}^{\bar{3}}\left\{ \bar{u}\bar{d}\right\} _{1}^{3}\right\rangle _{1}$  & 3167&	 876&	 1012&	-761&	11\\
\multirow{3}{*}{$(0)~1^{+}$}
 & $\left| \left(cs\right)_{1}^{\bar{3}}\left[\bar{u}\bar{d}\right]_{0}^{3}\right\rangle _{1}    $  & 3026&	 1002&	 949&	-815&	-139\\
 & $\left| \left(cs\right)_{0}^{6}\left[\bar{u}\bar{d}\right]_{1}^{\bar{6}}\right\rangle _{1}    $  & 3124&	 831&	 1015&	-752&	-6\\
 & $\left| \left(cs\right)_{1}^{6}\left[\bar{u}\bar{d}\right]_{1}^{\bar{6}}\right\rangle _{1}    $  & 3023&	 1028&	 912&	-838&	-114\\
\hline
\multirow{1}{*}{$(1)~2^{+}$}
 & $\left| \left(cs\right)_{1}^{\bar{3}}\left\{ \bar{u}\bar{d}\right\} _{1}^{3}\right\rangle _{2}$  & 3226&	 798&	 1060&	-726&	65\\
\multirow{1}{*}{$(0)~2^{+}$}
 & $\left| \left(cs\right)_{1}^{6}\left[\bar{u}\bar{d}\right]_{1}^{\bar{6}}\right\rangle _{2}    $  & 3188&	 805&	 1030&	-741&	57\\
\hline\hline
\end{tabular}
\end{center}
\end{table}

\begin{table*}
\begin{center}
\caption{Predicted mass spectra of $1S$ states for the $cn\bar{s}\bar{n}$ and $cs\bar{n}\bar{n}$ systems.}\label{abc}
\begin{tabular}{ccccccccccc}
\hline\hline
 & \multicolumn{4}{c}{$\underline{~~~~~~~~~~~~~~~~~~~~~~~~~~~~~~~~~~~~~~~~cn\bar{s}\bar{n}~~~~~~~~~~~~~~~~~~~~~~~~~~~~~~~~~~~~~~~~}$} & ~~~~  & \multicolumn{4}{c}{$\underline{~~~~~~~~~~~~~~~~~~~~~~~~~~~~~~~~~~~~~~~~cs\bar{n}\bar{n}~~~~~~~~~~~~~~~~~~~~~~~~~~~~~~~~~~~~~~~~}$} & \\
 & $(I)J^P$ & Configuration  & Eigenvector  & Mass (\text{MeV})  & ~~~~  & $(I)J^P$ & Configuration  & Eigenvector  & Mass (\text{MeV})  & \\
\hline
 & \multirow{2}{*}{$(0,1)~0^{+}$} & $\begin{array}{l}
\left|\left(cn\right)_{0}^{6}\left\{ \bar{s}\bar{n}\right\} _{0}^{\bar{6}}\right\rangle _{0}\\
\left|\left(cn\right)_{1}^{\bar{3}}\left\{ \bar{s}\bar{n}\right\} _{1}^{3}\right\rangle _{0}
\end{array}$  & $\left(\begin{array}{cc}
-0.65 & -0.76\\
-0.76 & 0.65
\end{array}\right)$  & $\left(\begin{array}{c}
3046\\
3279
\end{array}\right)$  & ~~~~  & $(1)~0^{+}$ & $\begin{array}{l}
\left|\left(cs\right)_{0}^{6}\left\{ \bar{u}\bar{d}\right\} _{0}^{\bar{6}}\right\rangle _{0}\\
\left|\left(cs\right)_{1}^{\bar{3}}\left\{ \bar{u}\bar{d}\right\} _{1}^{3}\right\rangle _{0}
\end{array}$  & $\left(\begin{array}{cc}
-0.65 & -0.76\\
-0.76 & 0.65
\end{array}\right)$  & $\left(\begin{array}{c}
3046\\
3260
\end{array}\right)$  & \\
 &  & $\begin{array}{l}
\left|\left(cn\right)_{0}^{\bar{3}}\left[\bar{s}\bar{n}\right]_{0}^{3}\right\rangle _{0}\\
\left|\left(cn\right)_{1}^{6}\left[\bar{s}\bar{n}\right]_{1}^{\bar{6}}\right\rangle _{0}
\end{array}$  & $\left(\begin{array}{cc}
-0.55 & -0.84\\
-0.84 & 0.55
\end{array}\right)$  & $\left(\begin{array}{c}
2828\\
3150
\end{array}\right)$  & ~~~~  & $(0)~0^{+}$ & $\begin{array}{l}
\left|\left(cs\right)_{0}^{\bar{3}}\left[\bar{u}\bar{d}\right]_{0}^{3}\right\rangle _{0}\\
\left|\left(cs\right)_{1}^{6}\left[\bar{u}\bar{d}\right]_{1}^{\bar{6}}\right\rangle _{0}
\end{array}$  & $\left(\begin{array}{cc}
-0.61 & -0.79\\
-0.79 & 0.61
\end{array}\right)$  & $\left(\begin{array}{c}
2818\\
3095
\end{array}\right)$  & \\
 & \multirow{2}{*}{$(0,1)~1^{+}$} & $\begin{array}{l}
\left|\left(cn\right)_{1}^{6}\left\{ \bar{s}\bar{n}\right\} _{0}^{\bar{6}}\right\rangle _{1}\\
\left|\left(cn\right)_{0}^{\bar{3}}\left\{ \bar{s}\bar{n}\right\} _{1}^{3}\right\rangle _{1}\\
\left|\left(cn\right)_{1}^{\bar{3}}\left\{ \bar{s}\bar{n}\right\} _{1}^{3}\right\rangle _{1}
\end{array}$  & $\left(\begin{array}{ccc}
0.66 & -0.59 & 0.46\\
-0.04 & 0.58 & 0.81\\
-0.75 & -0.56 & 0.36
\end{array}\right)$  & $\left(\begin{array}{c}
3067\\
3201\\
3245
\end{array}\right)$  & ~~~~  & $(1)~1^{+}$ & $\begin{array}{l}
\left|\left(cs\right)_{1}^{6}\left\{ \bar{u}\bar{d}\right\} _{0}^{\bar{6}}\right\rangle _{1}\\
\left|\left(cs\right)_{0}^{\bar{3}}\left\{ \bar{u}\bar{d}\right\} _{1}^{3}\right\rangle _{1}\\
\left|\left(cs\right)_{1}^{\bar{3}}\left\{ \bar{u}\bar{d}\right\} _{1}^{3}\right\rangle _{1}
\end{array}$  & $\left(\begin{array}{ccc}
0.67 & -0.59 & 0.44\\
-0.12 & 0.50 & 0.86\\
0.73 & 0.63 & -0.27
\end{array}\right)$  & $\left(\begin{array}{c}
3079\\
3185\\
3227
\end{array}\right)$  & \\
 &  & $\begin{array}{l}
\left|\left(cn\right)_{1}^{\bar{3}}\left[\bar{s}\bar{n}\right]_{0}^{3}\right\rangle _{1}\\
\left|\left(cn\right)_{0}^{6}\left[\bar{s}\bar{n}\right]_{1}^{\bar{6}}\right\rangle _{1}\\
\left|\left(cn\right)_{1}^{6}\left[\bar{s}\bar{n}\right]_{1}^{\bar{6}}\right\rangle _{1}
\end{array}$  & $\left(\begin{array}{ccc}
0.51 & -0.44 & 0.73\\
-0.70 & 0.27 & 0.66\\
0.49 & 0.85 & 0.17
\end{array}\right)$  & $\left(\begin{array}{c}
2949\\
3119\\
3199
\end{array}\right)$  & ~~~~  & $(0)~1^{+}$ & $\begin{array}{l}
\left|\left(cs\right)_{1}^{\bar{3}}\left[\bar{u}\bar{d}\right]_{0}^{3}\right\rangle _{1}\\
\left|\left(cs\right)_{0}^{6}\left[\bar{u}\bar{d}\right]_{1}^{\bar{6}}\right\rangle _{1}\\
\left|\left(cs\right)_{1}^{6}\left[\bar{u}\bar{d}\right]_{1}^{\bar{6}}\right\rangle _{1}
\end{array}$  & $\left(\begin{array}{ccc}
-0.66 & 0.41 & -0.63\\
-0.65 & 0.10 & 0.75\\
-0.37 & -0.91 & -0.20
\end{array}\right)$  & $\left(\begin{array}{c}
2946\\
3067\\
3161
\end{array}\right)$  & \\
 & \multirow{2}{*}{$(0,1)~2^{+}$} & $\left|\left(cn\right)_{1}^{\bar{3}}\left\{ \bar{s}\bar{n}\right\} _{1}^{3}\right\rangle _{2}$  & $\left(1\right)$  & $\left(3239\right)$  & ~~~~  & $(1)~2^{+}$ & $\left|\left(cs\right)_{1}^{\bar{3}}\left\{ \bar{u}\bar{d}\right\} _{1}^{3}\right\rangle _{2}$  & $\left(1\right)$  & $\left(3226\right)$  & \\
 &  & $\left|\left(cn\right)_{1}^{6}\left[\bar{s}\bar{n}\right]_{1}^{\bar{6}}\right\rangle _{2}$  & $\left(1\right)$  & $\left(3214\right)$  & ~~~~  & $(0)~2^{+}$ & $\left|\left(cs\right)_{1}^{6}\left[\bar{u}\bar{d}\right]_{1}^{\bar{6}}\right\rangle _{2}$  & $\left(1\right)$  & $\left(3188\right)$  & \\
\hline\hline
\end{tabular}
\end{center}
\end{table*}

\begin{figure}[htbp]
\begin{center}
\centering  \epsfxsize=9cm \epsfbox{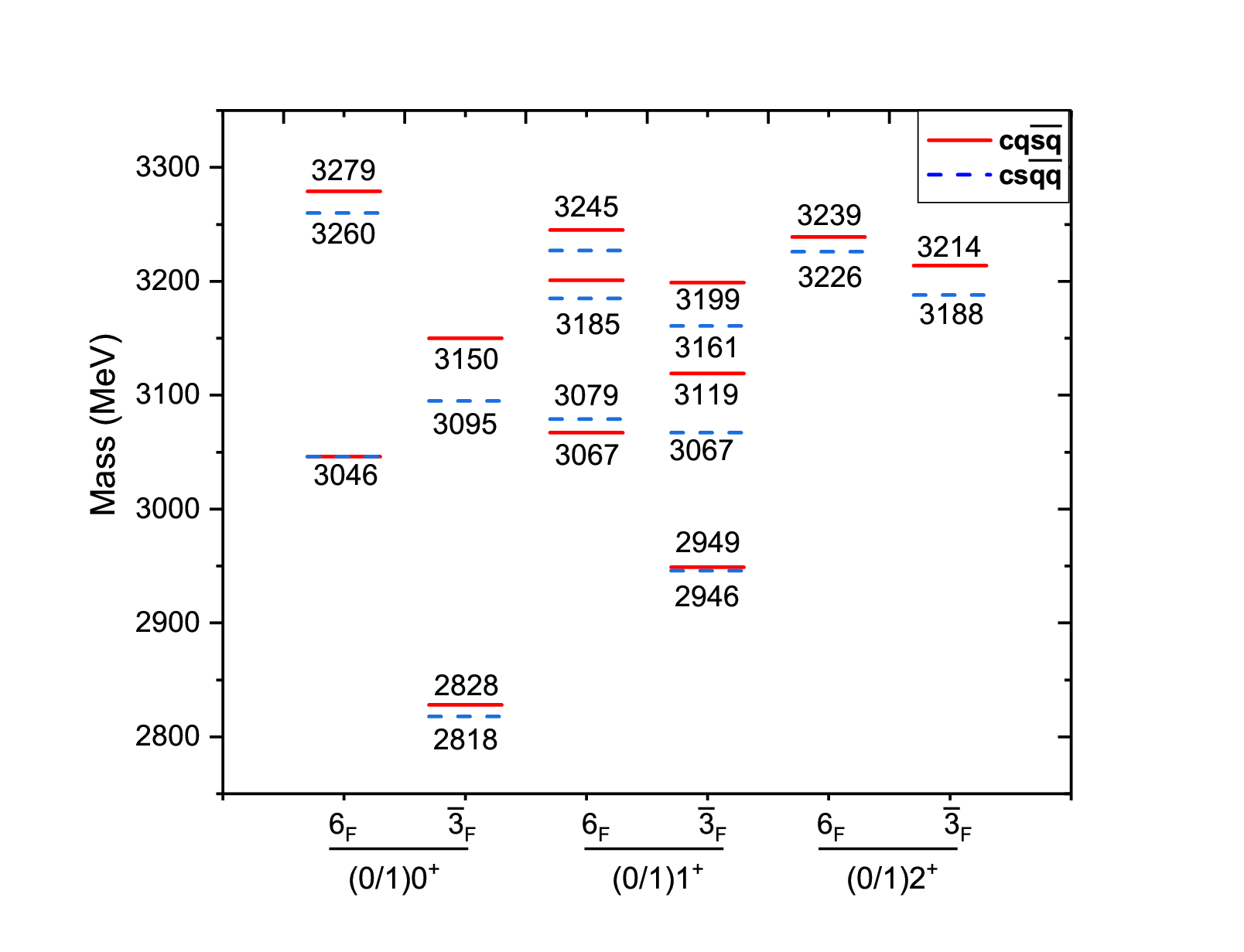}
\vspace{-1 cm}\caption{Mass spectra of $1S$-wave states for
the $cn\bar{s}\bar{n}$ (solid lines) and $cs\bar{n}\bar{n}$ (dashed lines) systems.} \label{spectrum}
\end{center}
\end{figure}

\section{Results and discussions}\label{result}

%\subsection{$cs\bar{q}\bar{q}$ system}\label{DISSCUS}

The predicted masses of each configuration for the $cs\bar{n}\bar{n}$ and $cn\bar{s}\bar{n}$ systems have been listed in Table~\ref{abcd}.
The contributions from each part of the Hamiltonian to these configurations are further analyzed. The results are listed in Table~\ref{abcd} as well. It shows that the averaged kinetic energy $\langle T\rangle$, the linear confining potential $\langle V^{Lin}\rangle$, and the Coulomb potential $\langle V^{Coul}\rangle$ have the same order of magnitude. Furthermore, it is found that
the spin-spin interaction plays an important role in some configurations belonging to $\bar{3}_F$:
$|(cn)_{0}^{\bar{3}}[\bar{s}\bar{n}]_{0}^{3}\rangle_{0} $/$|(cs)_{0}^{\bar{3}}[\bar{u}\bar{d}]_{0}^{3}\rangle_{0} $, $|(cn)_{1}^{6}[\bar{s}\bar{n}]_{1}^{\bar{6}}\rangle _{0}$/$|(cs)_{1}^{6}[\bar{u}\bar{d}]_{1}^{\bar{6}}\rangle _{0}$,
$|(cn)_{1}^{\bar{3}}[\bar{s}\bar{n}]_{0}^{3}\rangle_{1}$/$|(cs)_{1}^{\bar{3}}[\bar{u}\bar{d}]_{0}^{3}\rangle_{1}$, and $|(cn)_{1}^{6}[\bar{s}\bar{n}]_{1}^{\bar{6}}\rangle_{1}$/$|(cs)_{1}^{6}[\bar{u}\bar{d}]_{1}^{\bar{6}}\rangle_{1}$.
The predicted masses for these configurations are notably ($\sim100-200$ MeV) smaller than the other configurations due to the strong attractive spin-spin interactions $\langle V^{SS}\rangle\simeq -(100-200)$ MeV.

After considering configuration mixing, one can obtain the physical states.
The predicted mass spectrum for the $cs\bar{n}\bar{n}$ and $cn\bar{s}\bar{n}$ systems
have been given in Table~\ref{abc} and also shown in Fig.~\ref{spectrum}.
For the physical states with $J^P=0^+$ and $1^+$, there is strong mixing between different color configurations.
The configuration mixing effects can cause notable mass shifts to the physical states.
For example, the two $0^+$ configurations $| (cu)_{0}^{6}\{ \bar{s}\bar{d}\} _{0}^{\bar{6}}\rangle _{0}$ and
$| (cu)_{1}^{\bar{3}}\{ \bar{s}\bar{d}\} _{1}^{3}\rangle _{0}$  have comparable masses 3181 and 3145 MeV, respectively.
However, when including the configuration mixing effects the physical masses of the two $0^+$ states are
shifted to 3046 and 3279 MeV, respectively, the mass splitting can reach up to $\sim 230$ MeV.
It should be mentioned that the
$cn\bar{s}\bar{n}$ spectrum is slightly different from the $cs\bar{n}\bar{n}$ spectrum (see Fig.~\ref{spectrum}).
This difference comes from a slight SU(3) breaking effect of the $\bar{s}\bar{n}$ system considered
in our calculations.

\begin{table}
\begin{center}
\caption{The predicted decay widths $\Gamma$ (MeV) of the rearrangement decay processes of the ground $cs\bar{n}\bar{n}$ system. $T^a_{cs}$ and $T^f_{cs}$ stand for the states with $I=1$ and $I=0$, respectively.}\label{csqq}
\begin{tabular}{cc|cccc|c}\hline\hline																																	 
$SU(3)_F$ 	&	 State																															
	&	 $\Gamma_{T\rightarrow DK             }$																															 
	&	 $\Gamma_{T\rightarrow D^{*}K         }$																															 
	&	 $\Gamma_{T\rightarrow DK^{*}         }$																															 
	&	 $\Gamma_{T\rightarrow D^{*}K^{*}     }$																															 
	&	 $\Gamma_{sum}                               $ \\																															 
\hline																																	
\multirow{6}{*}{$6_{F}$}																																	 
 	&	 $T^{a}_{cs0}\left(3046\right)$ 	&	13.86 			&	   $\cdot\cdot\cdot$  			&	   $\cdot\cdot\cdot$  	&	8.82 	&									 22.68 	\\								
 	&	 $T^{a}_{cs0}\left(3260\right)$ 	&	0.18 			&	   $\cdot\cdot\cdot$  			&	   $\cdot\cdot\cdot$  	&	20.40 	&									 20.57 	\\								
 	&	 $T^{a}_{cs1}\left(3079\right)$ 	&	   $\cdot\cdot\cdot$  			 &	3.55 			&	2.14 	&	0.50 	&									 6.18 	\\								
 	&	 $T^{a}_{cs1}\left(3185\right)$ 	&	   $\cdot\cdot\cdot$  			 &	2.82 			&	0.69 	&	1.92 	&									 5.43 	\\								
 	&	 $T^{a}_{cs1}\left(3227\right)$ 	&	   $\cdot\cdot\cdot$  			 &	1.43 			&	7.06 	&	1.61 	&									 10.09 	\\								
 	&	 $T^{a}_{cs2}\left(3226\right)$ 	&	   $\cdot\cdot\cdot$  			 &	   $\cdot\cdot\cdot$  			&	   $\cdot\cdot\cdot$  	&	 2.74 	&									 2.74 	\\								
 \hline																																	
\multirow{6}{*}{$\bar{3}_{F}$}																																	 
 	&	 $[T^{f}_{cs0}\left(2818\right)]$ 	&	55.91 			&	   $\cdot\cdot\cdot$  			&	   $\cdot\cdot\cdot$  	&	   $\cdot\cdot\cdot$  	&									55.91 	\\								
 	&	 $\mathbf{T^{f}_{cs0}\left(2866\right)}$  	&	54.06 			&	   $\cdot\cdot\cdot$  			&	   $\cdot\cdot\cdot$  	&	   $\cdot\cdot\cdot$  	&									54.06 	\\								
 	&	 $T^{f}_{cs0}\left(3095\right)$ 	&	12.42 			&	   $\cdot\cdot\cdot$  			&	   $\cdot\cdot\cdot$  	&	102.68 	&									 115.10 	\\								
 	&	 $T^{f}_{cs1}\left(2946\right)$ 	&	   $\cdot\cdot\cdot$  			 &	33.86 			&	0.71 	&	2.88 	&									 37.46 	\\								
 	&	 $T^{f}_{cs1}\left(3067\right)$ 	&	   $\cdot\cdot\cdot$  			 &	2.71 			&	29.81 	&	20.00 	&									 52.52 	\\								
 	&	 $T^{f}_{cs1}\left(3161\right)$ 	&	   $\cdot\cdot\cdot$  			 &	4.91 			&	2.27 	&	13.02 	&									 20.20 	\\								
 	&	 $T^{f}_{cs2}\left(3188\right)$ 	&	   $\cdot\cdot\cdot$  			 &	   $\cdot\cdot\cdot$  			&	   $\cdot\cdot\cdot$  	&	 1.54 	&									 1.54 	\\								
\hline\hline																																	
\end{tabular}		
\end{center}
\end{table}

The rearrangement decay properties for the $cs\bar{n}\bar{n}$ and $cn\bar{s}\bar{n}$ systems have been
given in Tables~\ref{csqq} and~\ref{cqsq}, respectively. For the $cn\bar{s}\bar{n}$ system, we denote the tetraquark states by $T^a_{c\bar{s}}$ and $T^f_{c\bar{s}}$ with the superscripts ``$a$" and ``$f$" labelling their isospin $I=1$ and $I=0$, respectively. There are some interesting features arising from the width calculations. It is found that all the states of $6_F$ have a relatively narrow width within the range of
$\sim 1-30$ MeV. While for the states of $\bar{3}_F$, except the state with $J^P=2^+$, they have a
width within the range of $\sim 20-100$ MeV. For the $0^+$ and $1^+$ states, the rearrangement decay
is mainly driven by the spin-spin interactions. The decay amplitude caused by the confinement potential
part $V_{ij}^{cof}=-\frac{3}{16}({\vlab}_i\cdot{\vlab}_j)\left( b_{ij}r_{ij}-\frac{4}{3}\frac{\alpha_{ij}}{r_{ij}}+C_0 \right)$ is negligibly small.
The reason is that the two terms of the decay amplitude,
$\langle BC |V_{12}^{cof}+V_{34}^{cof}| A\rangle$ and $\langle BC |V_{14}^{cof}+V_{23}^{cof}| A\rangle$,
almost completely cancel out each other. The opposite signs of these two terms
come from the color factors. For the $2^+$ states, the rearrangement decay
is only driven by the confinement potential part, since the decay amplitudes
induced by the spin-spin interactions are zero.
One notices that all the $J^P=2^+$ states for both $I=0$ and $I=1$ are rather narrow and with a width of a few MeV.
The very narrow width nature is due to the strong cancelation between the two terms $\langle BC |V_{12}^{cof}+V_{34}^{cof}| A\rangle$
and $\langle BC |V_{14}^{cof}+V_{23}^{cof}| A\rangle$.

The new spin-0 state $X_0(2900)$ observed in the $D^-K^+$ channel at LHCb may be assigned as $\bar{T}_{cs0}^f(2818)$ with $I=0$, which is the antiparticle of $T_{cs0}^f(2818)$. It shows that both the predicted mass and spin-parity quantum numbers are consistent with the experimental observations. Taking the measured mass $M_{exp}=2866$ MeV for $\bar{T}_{cs0}^f(2818)$, the width is predicted to be $\Gamma\simeq 54$ MeV, which is in good agreement with the measured width $\Gamma_{exp}\simeq 57\pm 16$ MeV. With $X_0(2900)$
in the $D^-K^+$ channel assigned as the $\bar{T}_{cs0}^f(2818)$ state, a narrower state
$\bar{T}_{cs0}^a(3046)$ with a width of $\sim20$ MeV and a broader state
$\bar{T}_{cs0}^f(3095)$ with a width of $\sim100$ MeV may be observed in the same channel depending on the experimental statistics.
Their decay rates of $\bar{T}_{cs0}^a(3046)\to D^-K^+$ and $\bar{T}_{cs0}^f(3095)\to D^-K^+$ are
estimated to be $\sim 60\%$ and $\sim 10\%$, respectively.

Note that in Table~\ref{csqq}, $T_{cs0}^a(3260)$ has a very tiny decay rate ($\sim 10^{-2}$)
into the $DK$ channel because in the decay amplitude there is a strong cancelation between two different color structures $6\bar{6}$ and $\bar{3}3$. However, it shows that $T_{cs0}^a(3260)$ dominantly decays into $D^*K^*$ which almost saturates its total decay width. This seems to be a unique feature arising from this tetraquark system that a relatively narrow state should appear above the dominant decay channel. Thus, searching for $T_{cs0}^a(3260)$ in the $D^*K^*$ may provide a direct test of the tetraquark scenario and make it distinguishable from the hadronic molecule scenario.

As shown in Table~\ref{cqsq}, the spin-0 states $T^a_{c\bar{s}0}(2900)^{++/0}$ newly observed in
the $D_s^{+}\pi^+/D_s^{+}\pi^-$ channels at LHCb may be assigned to
be $T_{c\bar{s}0}^a(2828)$ with $I=1$ in our calculations.
The predicted mass, spin-parity numbers, and isospin are consistent with the observations.
Taking the measured mass $M_{exp}=2892$ MeV for $T_{c\bar{s}0}^a(2828)$,
the width is predicted to be $\Gamma\simeq 40$ MeV, which is slightly smaller than
the measured width $\Gamma_{exp}\simeq 136\pm 34$ MeV.
Except for $T_{c\bar{s}0}^a(2828)$, the other three states, $T^{a}_{c\bar{s}0}\left(3046\right)$, $T^{a}_{c\bar{s}0}\left(3279\right)$ and $T^{a}_{c\bar{s}0}\left(3150\right)$ with the flavor of $cn\bar{s}\bar{n}$, can also decay into $D_s\pi$ channel.
However, the decay rate of $T^{a}_{c\bar{s}0}\left(3279\right)\to D_s\pi$ turns to
be highly suppressed due a large cancelation between the two color structures $6\bar{6}$ and $\bar{3}3$.
The other two $0^+$ states $T^{a}_{c\bar{s}0}\left(3150\right)$ and $T^{a}_{c\bar{s}0}\left(3046\right)$
have sizeable decay rates into the $D_s\pi$ channel, their branching fractions
may reach up to $\sim 6\%$ and $\sim 20\%$, respectively. These two
$0^+$ states are likely to be observed in the $D_s\pi$ channel in future experiments.

If $T^a_{c\bar{s}0}(2900)^{++}$ is indeed
a tetraquark state predicted in our quark model, it should be
observed in the $D^+K^+$ channel as well, the partial width ratio between $D^+K^+$ and $D_s^+\pi^+$
is predicted to be
\begin{eqnarray}
\frac{\Gamma [D^+K^+]}{\Gamma [D_s^+\pi^+] }\simeq 1.4,
\end{eqnarray}
which can be used to test the nature of $T^a_{c\bar{s}0}(2900)^{++}$ in future experiments.
For the missing isospin triplet $T^a_{c\bar{s}0}(2900)^{+}$, the ideal observation channel is $D^{0}K^+$.
Furthermore, as the isospin partner of $T^a_{c\bar{s}0}(2900)^{+}$, the $I=0$ state $T_{c\bar{s}0}^f(2900)^+$
mainly decays into $D_s^+\eta$, and $D^0K^+/D^+K^0$ channels. This state may have potentials to be observed in $D^0K^+$ as well.
It should be mentioned that the $T^a_{c\bar{s}0}(2900)^{+}$ state may be broader than
$T^a_{c\bar{s}0}(2900)^{++,0}$, because the width of $T^a_{c\bar{s}0}(2900)^{+}$ may be enhanced by
the decay mechanism via $u\bar{u}$/$d\bar{d}$ annihilations~\cite{Anwar:2017toa}.

According to the rearrangement decay properties shown in Tables~\ref{csqq} and~\ref{cqsq},
more tetraquark states are expected to be observed in future experiments.
For the $cs\bar{u}\bar{d}$ system, two $I=1$ states $T^a_{cs1}(3079)^{0}$ and $T^a_{cs1}(3185)^{0}$, and two $I=0$ states $T^f_{cs1}(2946)^{0}$
and $T^f_{cs1}(3161)^{0}$ are most likely to be discovered in the $D^{*+}K^-$ channel; while the $I=1$ state $T^a_{cs1}(3227)^{0}$ and the $I=0$ state $T^f_{cs1}(3067)^{0}$ may have potentials to be found in the $D^0\bar{K}^{*0}$ final states.

For the $cn\bar{s}\bar{n}$ system with $I=1$, the medium width states $T^a_{c\bar{s}1}(2949)^{++}$ and $T^a_{c\bar{s}1}(2949)^{+}$
are most likely to be observed in the channels $D^{*+}K^+$ and $D^{*0}K^+$, respectively;
while the $T^a_{c\bar{s}1}(3067,3245)^{+}$ (belonging to $6_F$) and $T^a_{c\bar{s}1}(3119)^{+}$ (belonging to $\bar{3}_F$)
may have potentials to be found in the $D_s^+\rho^0$ final state.

For the $cn\bar{s}\bar{n}$ system with $I=0$, the $T^f_{c\bar{s}1}(3067,3245)^{+}$ (belonging to $6_F$) and
$T^f_{c\bar{s}1}(3119)^{+}$ (belonging to $\bar{3}_F$) may have potentials to be found in the
$D^+K^{*0}$ final states. They may highly overlap with their isospin  partners
$T^a_{c\bar{s}1}(3067)^{+}$ $T^a_{c\bar{s}1}(3119)^{+}$ and $T^a_{c\bar{s}1}(3245)^{+}$.

%{\color{red} *** The partial decay widths into the $D^*K^*$ channel for 2900 is sensitive to the phase space. It is better to extract the effective coupling of which a large value will be a strong indication for its molecular structure after the wavefunction renormalization.}

The wave functions obtained in this framework allow us to calculate the hadronic couplings for an initial tetraquark state to the final states. This is particularly interesting for those near-threshold states since the coupling strength can provide an indication of its structure arising from the near-threshold dynamics. Given that the partial decay width would generally be suppressed by the limited phase space, the effective coupling should be more useful for understanding the properties of the threshold states. Note that the predicted masses of some tetraquark states, such as
$T^{f}_{cs0}\left(2818\right)$, $T^{a,f}_{c\bar{s}0}\left(2828\right)$ and $T^{f}_{cs1}\left(2946\right)$,
are close to the mass threshold of the $D^*K^*$, $D_s^{*}\rho$ and $D_s^{*}\omega$ channels. The effective coupling constants for their couplings to the nearby $S$-wave thresholds should be more interesting.

To see the coupling strength for these $J^P=0^+,1^+$ tetraquark states
with $D^*K^*$, $D_s^{*}\rho$ and $D_s^{*}\omega$, we extracted the effective coupling constants defined by the following effective Lagrangians in terms of the quark model formalism, i.e.
\begin{equation}
\mathcal{L}_{SVV}=g_{SVV}\sqrt{m_{V_1}m_{V_2}}V_1^{\mu}V_{2\mu}\psi_S
\end{equation}
for the $0^+$ state coupling to $VV$, and
\begin{equation}
\mathcal{L}_{AVV}=g_{AVV}\epsilon^{\mu\nu\alpha\beta}
\partial_{\mu}V_{A\nu}V_{1\alpha}V_{2\beta}
\end{equation}
for the $1^+$ coupling to $VV$, where $V_1$ and $V_2$ stand for the vector meson fields;
$\psi_S$ and $V_A$ stand for the scalar and axial vector tetraquark fields, respectively.
$m_{V_1}$ and $m_{V_2}$ are masses of the vector mesons $V_1$ and $V_2$, respectively.

The results are listed in Table~\ref{coup}. It is found that the $0^+$ $T_{cs}$ state
with $I=0$, $T^{f}_{cs0}\left(3095\right)$, and two $0^+$ $T_{cs}$ states
with $I=1$, $T^{a}_{cs0}\left(3046\right)$ and $T^{a}_{cs0}\left(3260\right)$, may strongly
couple to the $D^*K^*$ channel. The three $0^+$ $T_{c\bar{s}}$ states
with $I=1$, $T^{a}_{c\bar{s}0}\left(3046\right)$, $T^{a}_{c\bar{s}0}\left(3279\right)$ and
$T^{a}_{c\bar{s}0}\left(3150\right)$, may strongly
couple to both $D^*K^*$ and $D_s^*\rho$ channels. Meanwhile, the three $0^+$ $T_{c\bar{s}}$ states
with $I=0$, $T^{f}_{c\bar{s}0}\left(3046\right)$, $T^{f}_{c\bar{s}0}\left(3279\right)$
and $T^{f}_{c\bar{s}0}\left(3150\right)$, are found to strongly
couple to both $D^*K^*$ and $D_s^*\omega$ channels. These strong couplings suggest that the final state interactions between these thresholds will be significantly enhanced by the tetraquark states. Following the picture of wave function renormalization for the final state interactions, it is possible that some of these tetraquark states should have a sizeable hadronic molecular component as the long-range part of the wave function~\cite{Weinberg:1962hj,Weinberg:1963zza,Weinberg:1965zz,Guo:2017jvc,Guo:2014iya}. Meanwhile, the short-range component should be driven by the color interactions among the constituent quarks as a tetraquark. While this scenario needs more elaborate investigations, we leave the systematic study of these issues in separate works in the future.

\begin{table*}
\begin{center}
\caption{The predicted decay widths $\Gamma$ (MeV) of the rearrangement decay processes of the ground $cn\bar{s}\bar{n}$ system. $T^a_{c\bar{s}}$ and $T^f_{c\bar{s}}$ stand for the states with $I=1$ and $I=0$, respectively.  }\label{cqsq}
\begin{tabular}{cc|ccccccccc}\hline\hline																																	 
% \multicolumn{11}{c}{decay properties of $6_F$ states}\\																																	
$SU(3)_F$ 	&	 State																															
	&	 $\Gamma_{T\to D_{s}\pi       }$																															 
	&	 $\Gamma_{T\to D_{s}^{*}\pi   }$																															 
	&	 $\Gamma_{T\to D_{s}\rho      }$																															 
	&	 $\Gamma_{T\to D_{s}^{*} \rho }$																															 
	&	 $\Gamma_{T\to DK             }$																															 
	&	 $\Gamma_{T\to D^{*}K         }$																															 
	&	 $\Gamma_{T\to DK^{*}         }$																															 
	&	 $\Gamma_{T\to D^{*}K^{*}     }$																															 
	&	 $\Gamma_{sum}                               $ \\																															 
\hline																																	
% 	&	                         	&	2103.00 			&	2247.00 			&	2738.00 	&	2882.00 	&	2359.00 	&	2502.00 	&	2757.00 	&	2900.00 	&	   \\%这一行是阈值。可以去掉									
%\hline																																	
\multirow{6}{*}{$6_{F}$}																																	 
 	&	 $T^{a}_{c\bar{s}0}\left(3046\right)$ 	&	5.41 			&	   $\cdot\cdot\cdot$  			&	   $\cdot\cdot\cdot$  	&	4.82 	&	 7.05 	&	   $\cdot\cdot\cdot$  	&	   $\cdot\cdot\cdot$  	&	10.79 	&	 28.07 	\\								
 	&	 $T^{a}_{c\bar{s}0}\left(3279\right)$ 	&	0.04 			&	   $\cdot\cdot\cdot$  			&	   $\cdot\cdot\cdot$  	&	7.72 	&	 0.03 	&	   $\cdot\cdot\cdot$  	&	   $\cdot\cdot\cdot$  	&	9.73 	&	 17.53 	\\								
 	&	 $T^{a}_{c\bar{s}1}\left(3067\right)$ 	&	   $\cdot\cdot\cdot$  			 &	1.12 			&	1.82 	&	0.35 	&	   $\cdot\cdot\cdot$  	&	 2.27 	&	4.36 	&	2.23 	&	12.15 	\\								
 	&	 $T^{a}_{c\bar{s}1}\left(3201\right)$ 	&	   $\cdot\cdot\cdot$  			 &	1.86 			&	0.01 	&	1.01 	&	   $\cdot\cdot\cdot$  	&	 1.78 	&	0.00 	&	2.40 	&	7.07 	\\								
 	&	 $T^{a}_{c\bar{s}1}\left(3245\right)$ 	&	   $\cdot\cdot\cdot$  			 &	0.43 			&	2.98 	&	0.02 	&	   $\cdot\cdot\cdot$  	&	 0.39 	&	3.73 	&	0.19 	&	7.75 	\\								
 	&	 $T^{a}_{c\bar{s}2}\left(3239\right)$ 	&	   $\cdot\cdot\cdot$  			 &	   $\cdot\cdot\cdot$  			&	   $\cdot\cdot\cdot$  	&	 0.56 	&	   $\cdot\cdot\cdot$  	&	   $\cdot\cdot\cdot$  	&	   $\cdot\cdot\cdot$  	 &	0.67 	&	1.24 	\\								
\hline																																	
 \multirow{6}{*}{$\bar{3}_{F}$}																																	 
 	&	 [$T^{a}_{c\bar{s}0}\left(2828\right)$] 	&	16.63 			&	   $\cdot\cdot\cdot$  			&	   $\cdot\cdot\cdot$  	&	   $\cdot\cdot\cdot$  	&	24.38 	&	   $\cdot\cdot\cdot$  	&	   $\cdot\cdot\cdot$  	&	   $\cdot\cdot\cdot$  	&	41.01 	\\								 
 	&	 $\mathbf{T^{a}_{c\bar{s}0}\left(2900\right)}^*$ 	&	16.11 			 &	   $\cdot\cdot\cdot$  			&	   $\cdot\cdot\cdot$  	&	 0.20 	&	 22.75 	&	   $\cdot\cdot\cdot$  	&	   $\cdot\cdot\cdot$  	&	0.86 	&	 39.91 	\\								
 	&	 $T^{a}_{c\bar{s}0}\left(3150\right)$ 	&	5.07 			&	   $\cdot\cdot\cdot$  			&	   $\cdot\cdot\cdot$  	&	34.51 	&	 5.97 	&	   $\cdot\cdot\cdot$  	&	   $\cdot\cdot\cdot$  	&	37.38 	&	 82.93 	\\								
 	&	 $T^{a}_{c\bar{s}1}\left(2949\right)$ 	&	   $\cdot\cdot\cdot$  			 &	6.86 			&	0.18 	&	0.00 	&	   $\cdot\cdot\cdot$  	&	 11.81 	&	1.15 	&	0.74 	&	20.75 	\\								
 	&	  $T^{a}_{c\bar{s}1}\left(3119\right)$ 	&	   $\cdot\cdot\cdot$  			 &	0.60 			&	6.82 	&	3.37 	&	   $\cdot\cdot\cdot$  	&	 1.18 	&	6.37 	&	4.42 	&	22.77 	\\								
 	&	  $T^{a}_{c\bar{s}1}\left(3199\right)$ 	&	   $\cdot\cdot\cdot$  			 &	1.99 			&	1.01 	&	4.77 	&	   $\cdot\cdot\cdot$  	&	 2.05 	&	1.36 	&	4.50 	&	15.69 	\\								
 	&	 $T^{a}_{c\bar{s}2}\left(3214\right)$ 	&	   $\cdot\cdot\cdot$  			 &	   $\cdot\cdot\cdot$  			&	   $\cdot\cdot\cdot$  	&	 0.45 	&	   $\cdot\cdot\cdot$  	&	   $\cdot\cdot\cdot$  	&	   $\cdot\cdot\cdot$  	 &	0.36 	&	0.81 	\\								
 \hline\hline																																	
 $SU(3)_F$ 	&	 State																															
	&	 $\Gamma_{T\to D_{s}\eta/D_{s}\eta'       }$																															 
	&	 $\Gamma_{T\to D_{s}^{*}\eta/D_{s}^{*}\eta'   }$																															 
	&	 $\Gamma_{T\to D_{s}\omega      }$																															 
	&	 $\Gamma_{T\to D_{s}^{*} \omega }$																															 
	&	 $\Gamma_{T\to DK             }$																															 
	&	 $\Gamma_{T\to D^{*}K         }$																															 
	&	 $\Gamma_{T\to DK^{*}         }$																															 
	&	 $\Gamma_{T\to D^{*}K^{*}     }$																															 
	&	 $\Gamma_{sum}                               $ \\																															 
\hline																																	
% 	&	                         	&	2103.00 			&	2247.00 			&	2738.00 	&	2882.00 	&	2359.00 	&	2502.00 	&	2757.00 	&	2900.00 	&	   \\%这一行是阈值。可以去掉									
%\hline																																	
\multirow{6}{*}{$6_{F}$}																																	 
 	&	 $T^{f}_{c\bar{s}0}\left(3046\right)$ 	&	4.18 	/	5.65 	&	   $\cdot\cdot\cdot$  			&	   $\cdot\cdot\cdot$  	&	4.81 	&	 7.05 	&	   $\cdot\cdot\cdot$  	&	   $\cdot\cdot\cdot$  	&	10.79 	&	 32.48 	\\								
 	&	 $T^{f}_{c\bar{s}0}\left(3279\right)$ 	&	0.04 	/	0.11 	&	   $\cdot\cdot\cdot$  			&	   $\cdot\cdot\cdot$  	&	7.82 	&	 0.03 	&	   $\cdot\cdot\cdot$  	&	   $\cdot\cdot\cdot$  	&	9.73 	&	 17.73 	\\								
 	&	 $T^{f}_{c\bar{s}1}\left(3067\right)$ 	&	   $\cdot\cdot\cdot$  			 &	1.01 	/	   $\cdot\cdot\cdot$  	&	1.84 	&	0.35 	&	   $\cdot\cdot\cdot$  	&	2.27 	&	4.36 	&	2.23 	&	12.06 	\\								
 	&	 $T^{f}_{c\bar{s}1}\left(3201\right)$ 	&	   $\cdot\cdot\cdot$  			 &	1.31 	/	1.51 	&	0.01 	&	1.03 	&	   $\cdot\cdot\cdot$  	&	 1.78 	&	0.00 	&	2.40 	&	8.04 	\\								
 	&	 $T^{f}_{c\bar{s}1}\left(3245\right)$ 	&	   $\cdot\cdot\cdot$  			 &	0.33 	/	0.51 	&	3.02 	&	0.02 	&	   $\cdot\cdot\cdot$  	&	 0.39 	&	3.73 	&	0.19 	&	8.20 	\\								
 	&	 $T^{f}_{c\bar{s}2}\left(3239\right)$ 	&	   $\cdot\cdot\cdot$  			 &	   $\cdot\cdot\cdot$  			&	   $\cdot\cdot\cdot$  	&	 0.58 	&	   $\cdot\cdot\cdot$  	&	   $\cdot\cdot\cdot$  	&	   $\cdot\cdot\cdot$  	 &	0.67 	&	1.25 	\\								
\hline																																	
 \multirow{6}{*}{$\bar{3}_{F}$}																																	 
 	&	 [$T^{f}_{c\bar{s}0}\left(2828\right)$] 	&	11.42 	/	   $\cdot\cdot\cdot$  	&	   $\cdot\cdot\cdot$  			&	   $\cdot\cdot\cdot$  	&	   $\cdot\cdot\cdot$  	&	24.38 	&	   $\cdot\cdot\cdot$  	&	   $\cdot\cdot\cdot$  	&	   $\cdot\cdot\cdot$  	 &	 35.80 	\\								
 	&	 $\mathbf{T^{f}_{c\bar{s}0}\left(2900\right)}^*$ 	&	11.46 	/	   $\cdot\cdot\cdot$  	&	   $\cdot\cdot\cdot$  			&	   $\cdot\cdot\cdot$  	&	0.16 	&	22.75 	&	   $\cdot\cdot\cdot$  	&	   $\cdot\cdot\cdot$  	&	0.86 	&	35.23 	\\								 
 	&	 $T^{f}_{c\bar{s}0}\left(3150\right)$ 	&	3.75 	/	5.54 	&	   $\cdot\cdot\cdot$  			&	   $\cdot\cdot\cdot$  	&	34.73 	&	 5.97 	&	   $\cdot\cdot\cdot$  	&	   $\cdot\cdot\cdot$  	&	37.38 	&	 87.37 	\\								
 	&	 $T^{f}_{c\bar{s}1}\left(2949\right)$ 	&	   $\cdot\cdot\cdot$  			 &	5.31 	/	   $\cdot\cdot\cdot$  	&	0.18 	&	0.00 	&	   $\cdot\cdot\cdot$  	&	11.81 	&	1.15 	&	0.74 	&	19.20 	\\								
 	&	  $T^{f}_{c\bar{s}1}\left(3119\right)$ 	&	   $\cdot\cdot\cdot$  			 &	0.62 	/	1.02 	&	6.91 	&	3.41 	&	   $\cdot\cdot\cdot$  	&	 1.18 	&	6.37 	&	4.42 	&	23.94 	\\								
 	&	  $T^{f}_{c\bar{s}1}\left(3199\right)$ 	&	   $\cdot\cdot\cdot$  			 &	1.54 	/	2.20 	&	1.03 	&	4.85 	&	   $\cdot\cdot\cdot$  	&	 2.05 	&	1.36 	&	4.50 	&	17.53 	\\								
 	&	 $T^{f}_{c\bar{s}2}\left(3214\right)$ 	&	   $\cdot\cdot\cdot$  			 &	   $\cdot\cdot\cdot$  			&	   $\cdot\cdot\cdot$  	&	 0.46 	&	   $\cdot\cdot\cdot$  	&	   $\cdot\cdot\cdot$  	&	   $\cdot\cdot\cdot$  	 &	0.36 	&	0.82 	\\								
 \hline\hline																																	
\end{tabular}																																	
\end{center}
\end{table*}

\begin{table*}
\begin{center}
\caption{The effective coupling constants for $T_{cs0,1}V_1V_2$ and $T_{c\bar{s}0,1} V_1V_2$.}\label{coup}
\begin{tabular}{cccccccccc}
\hline
\hline
 	&	 State  &	~~~ $g_{TD^{*}K^{*}}$ ~~~ 	&	 State  	&	~~~ $g_{TD_{s}^{*}\rho}$ ~~~	&	~~~  $g_{TD^{*}K^{*}}$ ~~~ 	&	 State  &	 $~~~g_{TD_{s}^{*}\omega}$ ~~~ 	&	 ~~~$g_{TD^{*}K^{*}}$~~~	 \\	
\hline
$6_{F}$~~~ &	 $T_{cs0}^{a}\left(3046\right)$  	&	1.35 	&	 $T_{c\bar{s}0}^{a}\left(3046\right)$  	&	1.00 	&	1.50 	&	 $T_{c\bar{s}0}^{f}\left(3046\right)$  	&	1.01 	&	1.50 	 \\	
 	&	 $T_{cs0}^{a}\left(3260\right)$  	&	1.61 	&	 $T_{c\bar{s}0}^{a}\left(3279\right)$  	&	0.99 	&	1.10 	&	 $T_{c\bar{s}0}^{f}\left(3279\right)$  	&	1.00 	&	1.10 	 \\	
 	&	 $T_{cs1}^{a}\left(3079\right)$  	&	0.11 	&	 $T_{c\bar{s}1}^{a}\left(3067\right)$  	&	0.09 	&	0.23 	&	 $T_{c\bar{s}1}^{f}\left(3067\right)$  	&	0.09 	&	0.23 	 \\	
 	&	 $T_{cs1}^{a}\left(3185\right)$  	&	0.18 	&	 $T_{c\bar{s}1}^{a}\left(3201\right)$  	&	0.13 	&	0.20 	&	 $T_{c\bar{s}1}^{f}\left(3201\right)$  	&	0.13 	&	0.20 	 \\	
 	&	 $T_{cs1}^{a}\left(3227\right)$  	&	0.16 	&	 $T_{c\bar{s}1}^{a}\left(3245\right)$  	&	0.02 	&	0.05 	&	 $T_{c\bar{s}1}^{f}\left(3245\right)$  	&	0.02 	&	0.05 	 \\	
\hline																		
$\bar{3}_{F}$~~~
 	&	 $T_{cs0}^{f}\left(2818\right)$  	&	   $0.49$
 	&	 $T_{c\bar{s}0}^{a}\left(2828\right)$  	&	0.32 	&	0.85 	&	 $T_{c\bar{s}0}^{f}\left(2828\right)$  	&	0.33 	&	0.85 	 \\	
 	&	 $T_{cs0}^{f}\left(3095\right)$  	&	4.04 	&	 $T_{c\bar{s}0}^{a}\left(3150\right)$  	&	2.24 	&	2.32 	&	 $T_{c\bar{s}0}^{f}\left(3150\right)$  	&	2.26 	&	2.32 	\\	
 	&	 $T_{cs1}^{f}\left(2946\right)$  	&	0.38 	&	 $T_{c\bar{s}1}^{a}\left(2949\right)$  	&	0.00 	&	0.19 	&	 $T_{c\bar{s}1}^{f}\left(2949\right)$  	&	0.00 	&	0.19 	 \\	
 	&	 $T_{cs1}^{f}\left(3067\right)$  	&	0.70 	&	 $T_{c\bar{s}1}^{a}\left(3119\right)$  	&	0.26 	&	0.30 	&	 $T_{c\bar{s}1}^{f}\left(3119\right)$  	&	0.26 	&	0.30 	 \\
 	&	 $T_{cs1}^{f}\left(3161\right)$  	&	0.49 	&	 $T_{c\bar{s}1}^{a}\left(3199\right)$  	&	0.28 	&	0.27 	&	 $T_{c\bar{s}1}^{f}\left(3199\right)$  	&	0.28 	&	0.27 	 \\	
\hline
\hline
\end{tabular}
\end{center}
\end{table*}

\section{summary}\label{summary}

In this work, we have studied the spectra of the $1S$-states for the $cs\bar{q}\bar{q}$ and $cq\bar{s}\bar{q}$ system within the NRPQM. To solve the four-body problem accurately, the explicitly correlated Gaussian method are adopted in our calculations. Furthermore, we have analyzed the rearrangement decays for the $1S$-states in a quark-exchange model by using the same integrations from the NRPQM.

Our studies show that most of the states lie in the mass range of $3.0-3.3$ GeV. For the states with $J^P=0^+$ and $1^+$, there is a strong mixing between different color configurations. Most of the states are narrow states with widths below $60$ MeV.
The decay amplitude caused by the confinement potential part is negligibly small due the strong
cancelations between the decay amplitudes $\langle BC |V_{12}^{cof}+V_{34}^{cof}| A\rangle$ and $\langle BC |V_{14}^{cof}+V_{23}^{cof}| A\rangle$.
The rearrangement decays of the tetraquarks may be mainly driven by the spin-spin interactions.
The decay amplitude caused by the confinement potential part is negligibly small.

Such a systematic phenomenon is useful for understanding the formation of relatively stable tetraquark states. The resonances, $X_0(2900)^0$ and $T^a_{c\bar{s}0}(2900)^{++/0}$ reported by LHCb can be assigned to be the lowest $1S$-wave tetraquark states $\bar{T}_{cs0}^f(2818)$ and $T^{a}_{c\bar{s}0}(2828)$, respectively. We also find that some of these near-threshold states have sizeable $S$-wave couplings to the corresponding open thresholds. This could be an indication for their hadronic molecule nature driven by the strong final state interactions via the tetraquark component. Based on our predictions, some of these $1S$-wave tetraquark states can be searched in other decay channels, such as $D^0K^+$, $D^+K^+$, $D^{*+}K^-$, $D^{*+}K^+$, $D^{*0}K^+$, $D^0\bar{K}^{*0}$, and $D_s^+\rho^0$, in future experiments.

\section*{Acknowledgements }

We would like to thank Mu-Yang Chen, Ming-Sheng Liu and Gang Li for valuable discussions.
This work is supported by the National Natural Science Foundation of China (Grants No.12175065, No.12235018, No.11775078, No.U1832173), 
and the Postgraduate Scientific Research Innovation Project of Hunan Province (Grant No. CX20220508).
Q.Z. is also supported in part, by the DFG and NSFC funds to the Sino-German CRC 110 ``Symmetries and the Emergence of Structure in QCD'' (NSFC Grant No. 12070131001, DFG Project-ID 196253076), National Key Basic Research Program of China under Contract No. 2020YFA0406300, and Strategic Priority Research Program of Chinese Academy of Sciences (Grant No. XDB34030302).

%\end{spacing}

\end{document}